\documentclass[aps,prb,twocolumn,superscriptaddress,showpacs]{revtex4}
\usepackage{psfrag,graphicx,stmaryrd,amsmath,amssymb}

\newcommand{\rv}{\mathbf{r}}
\newcommand{\kv}{\mathbf{k}}
\newcommand{\Rv}{\mathbf{R}}
\newcommand{\al}{\mathbf{a_{1}}}
\newcommand{\as}{\mathbf{a_{2}}}
\newcommand{\cv}{\mathbf{c}}
\newcommand{\dk}{\delta k}
\newcommand{\dkv}{\mathbf{\delta k}}
\newcommand{\K}{\mathbf{K}}

\begin{document}

\title{Local density of states and Friedel oscillations in graphene}

\author{\'Ad\'am B\'acsi}
\email[]{bacsi.adam@wigner.bme.hu}
\affiliation{Department of Physics, Budapest University of Technology and Economics, 1521 Budapest, Hungary}
\author{Attila Virosztek}
\email[]{viro@szfki.hu}
\affiliation{Department of Physics, Budapest University of Technology and Economics, 1521 Budapest, Hungary}
\affiliation{Research Institute for Solid State Physics and Optics, PO Box 49, 1525 Budapest, Hungary}

\date{\today}

\begin{abstract}
We investigate the local density of states and Friedel oscillation in
graphene around a well localized impurity in Born approximation.
In our analytical calculations Green's function technique has been used
taking into account both the localized atomic wavefunctions in a
tight-binding scheme and the
corresponding symmetries of the lattice. As a result we obtained long
wavelength oscillations in the density of electrons with long range
behavior proportional to the inverse square of the distance from the
impurity. These leading oscillations are out of phase on
nearby lattice sites (in fact for an extended defect they cancel each other
within one unit cell), therefore a probe with resolution worse than a few
unit cells will experience only the next to leading inverse cube decay of
density oscillations even for a short range scatterer.
\end{abstract}


\maketitle

Ever since the first production of atomically thin carbon films\cite{novossci}, graphene continues to fascinate physicists of almost all walks of life.
This simple two dimensional system of carbon atoms arranged in a honeycomb lattice provides us with a number of interesting properties due mainly
to the massless Dirac nature of the dispersion relation of its electrons\cite{castronrmp}. Undoped graphene has its Fermi energy at the tip of the
Dirac cones and behaves as a zero gap semiconductor. Applying appropriate gate voltage leads to electron or hole pockets and metallic behavior with
Fermi wavenumber $k_F$ typically much smaller than the size of the Brillouin zone. 
Exciting potential applications of graphene include
for example carbon based planar electronic circuitry with the possibility of electrically reconfigurable wiring\cite{williamscm}, exploiting the
guiding effect of graphene $p$-$n$ junctions with negative refractive index\cite{csertiprl}. On the theoretical side perhaps the simplest quantity to be
considered is the change in the local density of states (LDOS) and the Friedel oscillation (FO) in the excess charge density due to a well localized impurity.
Results of these considerations have implications for
the LDOS in disordered graphene\cite{zieglerprb}, for the interaction between adatoms in graphene\cite{shytovprl}, or in case of magnetic
adatoms\cite{uchoaprl} for the corresponding RKKY interaction\cite{saremiprb}.

Early theoretical work on the LDOS and the resulting FO around an impurity in graphene\cite{cheifalko,cheiepj} predicted long
wavelength ($2k_F$) oscillations in the charge density, but with envelope decaying like $r^{-3}$ at distance $r$ from the impurity. This is in
contrast to the $r^{-2}$ decay in a degenerate nonrelativistic two dimensional Fermi gas, and suppressed backscattering of chiral graphene electrons
residing around the Fermi circle of the Dirac cone
was offered as an explanation. However, graphene has two inequivalent Dirac cones (valleys) in the Brillouin zone, and intervalley scattering
by the impurity may lead to short wavelength oscillations on the order of a few lattice constants as well. Indeed, a scanning tunneling microscopy
(STM) study\cite{ruttersci} of epitaxial
graphene revealed two different length scales around defects. Subsequently, intervalley (or internodal) scattering has been built in the
theory\cite{benaprl}, and has been shown to be responsible for $r^{-1}$ decay in LDOS and $r^{-2}$ decay in FO. The different power laws for intranodal
and internodal scattering have been observed by Fourier Transform STM (FT-STM) experiment\cite{brihuprl}, but no detailed investigation of the short
wavelength oscillations has been performed either experimentally or theoretically. The first step towards this direction has been made by incorporating
atomic wavefunctions instead of just plane waves into the theory\cite{benaprb}, and it has become clear that although the LDOS falls off with $r^{-1}$,
for intranodal scattering only it has opposite sign on the two sublattices. Therefore upon coarse-grainig, for example due to experimental resolution, the
leading order decay cancels within one unit cell, and one is left with the next to leading $r^{-2}$ envelope, and consequently an $r^{-3}$ decay of FO.

\begin{figure}[ht]
\includegraphics[width=66.5mm]{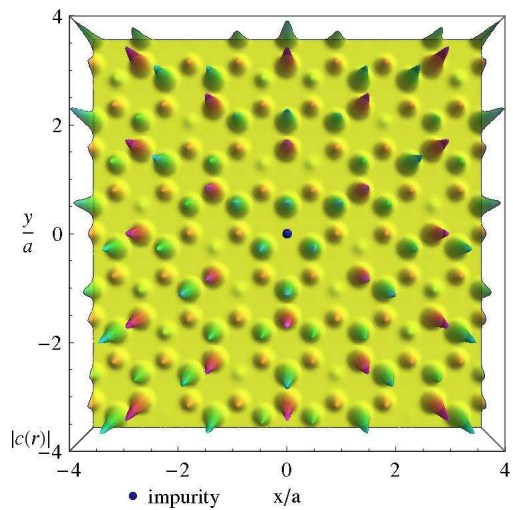}
\caption{Short wavelength spatial dependence of the LDOS due to an impurity at the center. 3D plot of $c_{A}(\rv)+c_{B}(\rv)=|c(\rv)|$ is viewed from above, and the color code
indicates the opposite sign on the two sublattices.\label{fig:n}}
\end{figure}

The aim of the present paper is to give a detailed analysis of the short wavelength oscillations due to internodal scattering within the framework of a
simple and transparent, atomic resolution theory based on the tight-binding wavefunctions of electrons on the honeycomb lattice. Our main results are the short
range spatial pattern of LDOS and FO on both sublattices with the corresponding symmetries (see Fig. \ref{fig:n}), and the observation that including 
internodal scattering still lead to cancellation of the leading power law decay of oscillations, but not within one unit cell, but within three 
neighboring ones. This means that if for example the STM tip can not resolve the six atoms
on a hexagon of the honeycomb lattice, we encounter again just the next to leading power law envelopes for both LDOS and FO.

We begin with the tight-binding Hamiltonian of carbon atoms forming a honeycomb lattice with nearest neighbor hopping $t$ (assumed to be real) between
the $p_z$ orbitals $\varphi(\rv)$.
In momentum space this leads to a 2x2 Hamiltonian matrix spanned by Bloch waves constructed from the atomic orbitals on the two sublattices A and B as

\begin{equation}
\label{eq:hamiltonian}
H=\sum_{\kv\sigma}
\left(\begin{array}{c c}
a_{A\kv\sigma}^{+} & a_{B\kv\sigma}^{+}
\end{array}\right)\left(
\begin{array}{c c}
0 & tf^{*}(\kv) \\
tf(\kv) & 0
\end{array}\right)\left(
\begin{array}{c}
a_{A\kv\sigma}\\
a_{B\kv\sigma}
\end{array}\right)\,,
\end{equation}
where $f(\kv)=1+e^{i\kv\al}+e^{i\kv\as}$ with primitive lattice vectors $\mathbf{a_{1,2}}=a(\pm 1/2,\sqrt{3}/2)$ of length $a$. Diagonalization yields the eigenvalues
$\varepsilon_{\pm}(\kv)=\pm|t||f(\kv)|$ defining the two bands of pure graphene, and the eigenvectors in the tight-binding form

\begin{eqnarray}
\label{eq:eigfun}
\Psi_{\pm,\kv}(\rv)=\frac{1}{\sqrt{N}}\sum_{\Rv}e^{i\kv\Rv} \Big[\mp e^{-i\delta(\kv)}\varphi(\rv-\Rv)+\nonumber\\
+\varphi(\rv-\Rv-\cv)\Big]\,,
\end{eqnarray}
where 
the summation runs over the $\Rv$ lattice vectors and 
$\delta(\kv)$ is the complex phase of $f(\kv)$ determining the mixing of A and B sublattice states, and $\cv=(\al+\as)/3$ is the vector pointing from the A
to the B site within the unit cell. As is well known, $f(\kv)$ vanishes at the corners of the hexagonal Brillouin zone, leading to the
massless Dirac spectrum $\varepsilon_{\pm}(\kv)\approx\pm\hbar v_{F}\dk\,,$ with $v_{F}=\sqrt{3}a|t|/2\hbar$ the Fermi velocity, and $\dkv$ the wavenumber
measured from the corners. Since there are two inequivalent corners, for example $\K=(\mathbf{b_1}-\mathbf{b_2})/3$ and $\K'=-\K$ as expressed by the primitive
reciprocal lattice vectors, there are two Dirac cones or valleys in the Brillouin zone. Electron spin does not play any role in the forthcoming discussion, therefore we
suppress spin indices, and all our results will refer to one spin orientation.

The real space representation of the Green's function of complex energy variable $G(z,\rv,\rv ')$ is of central importance for our subject, since the LDOS is given
by its analytic continuation just above the real axis as $\rho(\varepsilon,\rv)=-\pi^{-1}\mathrm{Im}G(\varepsilon+i\delta,\rv,\rv)$. For pure graphene, not perturbed by impurities, 
the free Green's function is evaluated using Eq.(\ref{eq:eigfun}) as

\begin{eqnarray}
\label{eq:greenfn}
G^{0}(z,\rv,\rv ')=\sum_{l=\pm,\kv}\frac{\Psi_{l,\kv}(\rv)\Psi_{l,\kv}^{*}(\rv ')}{z-\varepsilon_{l}(\kv)}=\nonumber\\
=\sum_{\Rv\Rv '}\underline{\varphi}(\rv-\Rv)\mathbf{G}^{0}(z,\Rv-\Rv ')\underline{\varphi}^{+}(\rv '-\Rv ')\,,
\end{eqnarray}
where $\underline{\varphi}(\rv)=[\varphi(\rv),\varphi(\rv-\cv)]$ is a row vector formed by the orbitals of the A and B sites in the unit cell, and the Green's
matrix is given by

\begin{equation}
\label{eq:greenmatrix}
\mathbf{G}^{0}(z,\Rv)=\frac{1}{N}\sum_{\kv}\frac{e^{i\kv\Rv}}{z^{2}-t^{2}|f(\kv)|^{2}}\left(\begin{array}{c c}
z & tf^{*}(\kv) \\
tf(\kv) & z
\end{array} \right)\,.
\end{equation}
The free Green's function in Eq.(\ref{eq:greenfn}) is clearly translation invariant by a lattice vector only, but due to the well localized nature of the atomic
wavefunctions it typically consists of only one non-negligible term determined by the proximity of $\rv$ and $\rv '$ to a lattice point given by either $\rv_A(=\Rv)$
on the A sublattice, or by $\rv_B(=\Rv+\cv)$ on the B sublattice. On the other hand, the (identical) diagonal elements of the Green's matrix in
Eq.(\ref{eq:greenmatrix}) are even
functions of $\Rv$ (eg. $G^0_{AA}(z,-\Rv)=G^0_{AA}(z,\Rv)$), while the off diagonal elements transform into each other upon reflection ($G^0_{AB}(z,-\Rv)=G^0_{BA}(z,\Rv)$).
It is instructive to consider the LDOS of the pure system given by
$\rho_0(\varepsilon,\rv)=\rho_0(\varepsilon)\sum_{\Rv}[|\varphi(\rv-\Rv)|^2+|\varphi(\rv-\Rv-\cv)|^2]/2$,
where $\rho_0(\varepsilon)$ is the DOS, which can be approximated around the Dirac point by $A_c|\varepsilon|/\pi(\hbar v_{F})^2$, $A_c=\sqrt{3}a^2/2$ being the area of the
unit cell. 

Let us consider a substitutional impurity characterized by a short range scattering potential energy $U(\rv)=u\delta(\rv)$ located at the origin, which is a site on the A
sublattice. The correction to the LDOS in Born approximation is determined by $\Delta G(z,\rv,\rv)=uG^{0}(z,\rv,0)G^{0}(z,0,\rv)$, which
can be expressed due to the localized atomic orbitals as

\begin{eqnarray}
\label{eq:gga}
 &u_0|\varphi(\rv-\rv_{A})|^2G^0_{AA}(z,\rv_A)G^0_{AA}(z,-\rv_A)\,,\quad\mbox{or}\\
\label{eq:ggb}
 &u_0|\varphi(\rv-\rv_B)|^2G^0_{BA}(z,\rv_B-\cv)G^0_{AB}(z,-\rv_B+\cv)\,,
\end{eqnarray}
depending on whether $\rv$ is in the vicinity of an A or a B site respectively, and $u_0=u|\varphi(0)|^2$. Clearly the spatial pattern is dominated by the density
profile of the atomic orbital centered on the given lattice site, while the multiplicative factor on that site depends on the corresponding Green's matrix elements.
It can be proven that
the diagonal elements of the Green's matrix like $G^0_{AA}(z,\rv_A)$ have sixfold rotational symmetry in $\rv_A$, while the off diagonal elements like
$G^0_{BA}(z,\rv_B-\cv)$ have only threefold rotational symmetry in $\rv_B$.

In order to evaluate the relevant Green's matrix elements from Eq.(\ref{eq:greenmatrix}) we observe that the most important contributions come from the nodal points of the
spectrum, i.e. from around the points $\K$ and $\K'=-\K$ in the Brillouin zone. The matrix elements therefore consist of two terms each, led by fast oscillations of the
type $\exp(i\K\Rv)$ and $\exp(i\K'\Rv)$, modulated by functions of slow spatial variation. These latter functions can be evaluated by using the linearized
$f(\pm\K+\dkv)=-\sqrt{3}a(\pm\dk_x+i\dk_y)/2$ expression around the nodal points up to a cutoff $k_c$. This will be a good approximation
for these slowly varying factors of the Green's matrix elements for distances from the impurity much larger, and for characteristic spatial variations much longer
than $1/k_c$. This procedure leads to the following result for the diagonal element in Eq.(\ref{eq:gga}):

\begin{equation}
\label{eq:ga}
G^0_{AA}(z,\rv_A)=(e^{i\K\rv_A}+e^{i\K'\rv_A})\frac{-iA_cz}{(2\hbar v_F)^2}H_{0}^{(1)}\left(\frac{z r_{A}}{\hbar v_{F}}\right)\,,
\end{equation}
where $H_{0}^{(1)}(z)$ is the Hankel function, and this functional form is valid for $\mathrm{Im}z>0$ only (which is enough for the prescribed analytic continuation)
and for $|z|\ll\hbar v_{F}k_c$. Similarly, for the off diagonal element in Eq.(\ref{eq:ggb}) we obtain
\begin{eqnarray}
\label{eq:gb}
G^0_{BA}(z,\rv_B-\cv)=(e^{i(\K\rv_B+\vartheta)}-e^{i(\K'\rv_B-\vartheta)})\times\nonumber\\
\times\frac{-\mathrm{sgn}(t)A_cz}{(2\hbar v_F)^2}H_{1}^{(1)}\left(\frac{z r_{B}}{\hbar v_{F}}\right)\,,
\end{eqnarray}
where $H_{1}^{(1)}(z)=-dH_{0}^{(1)}(z)/dz$, $\vartheta$ is the angle $\rv_B$ makes with the $x$ axis (which is parallel to $\K$). For
$G^0_{AB}(z,-\rv_B+\cv)$ we have the same formula as in Eq.(\ref{eq:gb}), except that the prefactor with the exponentials is replaced by
$(-e^{-i(\K\rv_B+\vartheta)}+e^{-i(\K'\rv_B-\vartheta)})$. We remind the reader that these exponential prefactors in the Green's matrix elements describing short wavelength
oscillations are exact, as opposed to the spatial dependence described by the Hankel functions.

The change of the LDOS in graphene due to the impurity $\Delta\rho(\varepsilon,\rv)=-\pi^{-1}\mathrm{Im}\Delta G(\varepsilon+i\delta,\rv,\rv)$ is now easily calculated using
Eqs.(\ref{eq:gga}-\ref{eq:gb}). Let us note first, that the short wavelength spatial pattern in eg. Eq.(\ref{eq:gga}) 
is given as $(e^{i\K\rv_{A}}+e^{i\K'\rv_{A}})(e^{-i\K\rv_{A}}+e^{-i\K'\rv_{A}})=4\cos^2(\K\rv_A)$. 
However,
if the impurity is unable to produce intervalley scattering
(eg. because it is extended and has no large wavenumber Fourier components), then plane waves belonging to different valleys will not contribute to the above product, and
we will only have $1+1=2$ as a result. Thus, without intervalley scattering, the factor $\cos^2(\K\rv_A)$ describing the short wavelength spatial pattern
should be replaced by its average i.e. $1/2$. The same is true for the factor $\sin^2(\K\rv_B+\vartheta)$ appearing in Eq.(\ref{eq:ggb}).

Returning to the change in the LDOS, after analytic continuation we obtain
\begin{equation}
\label{eq:dra}
\Delta\rho(\varepsilon,\rv)=\frac{\pi}{2}u_0c_A(\rv)\rho_0^2(\varepsilon)\mathrm{sgn}(\varepsilon)J_{0}\left(\frac{|\varepsilon|r_{A}}{\hbar v_{F}}\right)
Y_{0}\left(\frac{|\varepsilon|r_{A}}{\hbar v_{F}}\right)
\end{equation}
if $\rv$ is in the vicinity of an A site, and
\begin{equation}
\label{eq:drb}
\Delta\rho(\varepsilon,\rv)=\frac{\pi}{2}u_0c_B(\rv)\rho_0^2(\varepsilon)\mathrm{sgn}(\varepsilon)J_{1}\left(\frac{|\varepsilon|r_{B}}{\hbar v_{F}}\right)
Y_{1}\left(\frac{|\varepsilon|r_{B}}{\hbar v_{F}}\right)
\end{equation}
if $\rv$ is in the vicinity of a B site. Here $J_{0}$, $J_{1}$ and $Y_{0}$, $Y_{1}$ are Bessel functions of the first and the second kind. The factors
describing the short wavelength spatial behavior are given by
\begin{eqnarray}
\label{eq:ca}
 &c_{A}(\rv)=|\varphi(\rv-\rv_{A})|^{2}\cos^{2}(\K\rv_{A})\,,\, \mbox{and}\\
\label{eq:cb}
 &c_{B}(\rv)=|\varphi(\rv-\rv_{B})|^{2}\sin^{2}(\K\rv_{B}+\vartheta)\,.
\end{eqnarray}
It can be easily proven that the $c_{A}(\rv)$ and $c_{B}(\rv)$ factors are invariant under sixfold and threefold rotations respectively. 
Let us remember that for intravalley scattering only, the weights of the density of atomic orbitals would be
the same on each atoms.

The isotropic spatial dependence in the LDOS given by the Bessel functions in Eqs.(\ref{eq:dra},\ref{eq:drb}) describes long wavelength
oscillations, since the characteristic wavenumber $k=\varepsilon/\hbar v_F$ is much smaller than the cutoff $k_c$. Consequently, instead of $r_{A}$ and $r_{B}$ we can
use here the continous variable $r$ measuring the distance from the impurity. For distances large enough to satisfy $|k|r\gg 1$ we can use the leading asymptotic
expressions $J_1(x)Y_1(x)=\cos(2x)/\pi x=-J_0(x)Y_0(x)$ for $x\rightarrow\infty$. Due to this sign change on the two sublattices, it is useful to define a composit
short wavelength pattern $c(\rv)=-c_{A}(\rv)$ if $\rv$ is near an A site, and $c(\rv)=c_{B}(\rv)$ if $\rv$ is near a B site. Then the change of the LDOS due to an
impurity at the origin can be given by the following compact formula:
\begin{equation}
\label{eq:dr}
\Delta\rho(\varepsilon,\rv)=u_0c(\rv)\rho_0^2(\varepsilon)\frac{\cos(2kr)}{2kr}\,,
\end{equation}
indicating a long wavelength oscillation with an $r^{-1}$ decay and a short wavelength spatial pattern given by $c(\rv)$, and shown in Fig. \ref{fig:n}. Although a
very rich structure can be seen on the plot, the overall threefold rotational symmetry is apparent, and appears to have been observed experimentally\cite{simonepjb}.

Due to finite resolution however, the STM tip will measure a spatial average of the pattern on Fig. \ref{fig:n}. In the simpler case of an extended impurity not
producing intervalley scattering, $c(\rv)$ has atomic densities with equal weight on each site of one sublattice and the same weight with opposite sign on each site of
the other sublattice. Clearly, a resolution worse than an elementary cell of graphene will lead to cancellation of the leading $r^{-1}$ decay in Eq.(\ref{eq:dr}),
leaving us with the next to leading $r^{-2}$ decay of LDOS. On the other hand for a short range impurity potential, internodal scattering contributes as well, and 
yields the short wavelength spatial pattern on Fig. \ref{fig:n}, where no cancellation within one unit cell occurs. However it is easily shown, that averaging
$c(\rv)$ over three neighboring unit cells again leads to cancellation of the weights of the atomic densities. Indeed if the weights in $c(\rv)$ are added
in the unit cells given by lattice vectors $\Rv$, $\Rv+\al$ and $\Rv+\al-\as$ (far from the impurity $\vartheta$ is the same for all three lattice sites), we obtain zero. 
Therefore an STM with resolution worse than three elementary
cells will not be able to measure the leading $r^{-1}$ decay either. It is easily shown that the weights of the atomic densities cancel on the six sites of any
hexagon of the honeycomb lattice.

The change in the particle density at zero temperature due to the impurity is now easily obtained by integrating the LDOS up to
the Fermi energy $\varepsilon_{F}$.
Since $\Delta\rho(\varepsilon,\rv)$ is an odd function of energy, $\Delta n(\rv)$ is even in $\varepsilon_F$, therefore we integrate up to $-|\varepsilon_F|$. As long as
$k_Fr\gg 1$, where $k_F=|\varepsilon_F|/\hbar v_F$, we can use the asymptotic forms of the Bessel functions to obtain the leading term of the FO as

\begin{equation}
\label{eq:dn}
\Delta n(\rv)=\frac{u_0A_c}{4\pi}c(\rv)\rho_0(\varepsilon_F)\frac{\sin(2k_Fr)}{r^2}\,,
\end{equation}
where we disregarded the oscillations due to the cutoff at the lower limit of integration as unphysical, since these are too fast to be taken seriously in our scheme.
The FO with wavenumber $2k_F$ decays on both sublattice as $r^{-2}$, but as it was discussed in case of the LDOS, limited experimental resolution leads to cancellation
of the leading power laws, and only the $r^{-3}$ decay will be observed. For resolution of about a unit cell this happens for an extended impurity not producing
internodal scattering, but somewhat worse resolution of about three unit cells or a hexagon of the honeycomb lattice leads to cancellation even for a pointlike
impurity producing internodal scattering as well. Finally it is worth mentioning that for half filled graphene, i.e. $\varepsilon_F=0$, we obtain $r^{-3}$ decay without
$2k_F$ oscillations on both sublattices.

In conclusion, we have presented a clear and concise calculation of the change in the local density of states, and the resulting Friedel oscillations in the electron
density in graphene, due to a single nonmagnetic substitutional impurity with short range scattering potential. In order to achieve atomic scale description we used
tight-binding wavefunctions with atomic orbitals. As a consequence of this approach, the short wavelength spatial pattern in our results for both LDOS and FO, described
by $c(\rv)$ in Eqs.(\ref{eq:dr},\ref{eq:dn}), can be considered exact. Using a linearized electronic spectrum up to a cutoff affects only the rest of the spatial
dependence, but for energies close to the Dirac point and for distances far from the impurity, the long wavelength oscillating parts are excellent approximations as well.
In particular, the LDOS decays as $r^{-1}$ and the FO with wavenumber $2k_F$ as $r^{-2}$ on both sublattices. The short wavelength spatial pattern, depicted on Fig. \ref{fig:n}, 
shows the required rotational symmetries. Since $c(\rv)$ has alternating signs on neighboring lattice sites, experimental resolution worse than about three unit cells 
(or a hexagon of the lattice) will lead to cancellation of the leading power law decays in both quantities yielding the next to leading $r^{-2}$ and $r^{-3}$ behavior for 
LDOS and FO respectively. Within the present framework, the case of an extended defect can also be considered by restricting the calculation to intravalley scattering 
by the impurity. This affects only the short wavelength spatial pattern $c(\rv)$, making the weights of the atomic densities in it equal (but still of opposite sign on the two sublattices). 
Consequently, the above mentioned cancellation of the leading power law decays occurs already at resolution worse than one unit cell. Therefore if the experimental resolution 
falls between one and three unit cells, distinction can be made between localized and extended defects, since the observed power law decay for LDOS for example should 
follow $r^{-1}$ for the former and $r^{-2}$ for the latter.

\begin{acknowledgments}
We have benefited from discussions with B. D\'ora and J. Cserti. This work was supported by the Hungarian Scientific Research Fund under Grants No. OTKA K72613, 
and T\'AMOP-4.2.1/B-09/1/KMR-2010-0002.
\end{acknowledgments}

\bibliography{graphene}

\begin{thebibliography}{15}
\expandafter\ifx\csname natexlab\endcsname\relax\def\natexlab#1{#1}\fi
\expandafter\ifx\csname bibnamefont\endcsname\relax
  \def\bibnamefont#1{#1}\fi
\expandafter\ifx\csname bibfnamefont\endcsname\relax
  \def\bibfnamefont#1{#1}\fi
\expandafter\ifx\csname citenamefont\endcsname\relax
  \def\citenamefont#1{#1}\fi
\expandafter\ifx\csname url\endcsname\relax
  \def\url#1{\texttt{#1}}\fi
\expandafter\ifx\csname urlprefix\endcsname\relax\def\urlprefix{URL }\fi
\providecommand{\bibinfo}[2]{#2}
\providecommand{\eprint}[2][]{\url{#2}}

\bibitem[{\citenamefont{Novoselov et~al.}(2004)\citenamefont{Novoselov, Geim,
  Morozov, Jiang, Zhang, Dubonos, Grigorieva, and Firsov}}]{novossci}
\bibinfo{author}{\bibfnamefont{K.~S.} \bibnamefont{Novoselov}},
  \bibinfo{author}{\bibfnamefont{A.~K.} \bibnamefont{Geim}},
  \bibinfo{author}{\bibfnamefont{S.~V.} \bibnamefont{Morozov}},
  \bibinfo{author}{\bibfnamefont{D.}~\bibnamefont{Jiang}},
  \bibinfo{author}{\bibfnamefont{Y.}~\bibnamefont{Zhang}},
  \bibinfo{author}{\bibfnamefont{S.~V.} \bibnamefont{Dubonos}},
  \bibinfo{author}{\bibfnamefont{I.~V.} \bibnamefont{Grigorieva}},
  \bibnamefont{and} \bibinfo{author}{\bibfnamefont{A.~A.}
  \bibnamefont{Firsov}}, \bibinfo{journal}{Science}
  \textbf{\bibinfo{volume}{306}}, \bibinfo{pages}{666} (\bibinfo{year}{2004}).

\bibitem[{\citenamefont{{Castro Neto} et~al.}(2009)\citenamefont{{Castro Neto},
  Guinea, Peres, Novoselov, and Geim}}]{castronrmp}
\bibinfo{author}{\bibfnamefont{A.~H.} \bibnamefont{{Castro Neto}}},
  \bibinfo{author}{\bibfnamefont{F.}~\bibnamefont{Guinea}},
  \bibinfo{author}{\bibfnamefont{N.~M.~R.} \bibnamefont{Peres}},
  \bibinfo{author}{\bibfnamefont{K.~S.} \bibnamefont{Novoselov}},
  \bibnamefont{and} \bibinfo{author}{\bibfnamefont{A.~K.} \bibnamefont{Geim}},
  \bibinfo{journal}{Rev. Mod. Phys.} \textbf{\bibinfo{volume}{81}},
  \bibinfo{pages}{109} (\bibinfo{year}{2009}).

\bibitem[{\citenamefont{Williams et~al.}()\citenamefont{Williams, Low,
  Lundstrom, and Marcus}}]{williamscm}
\bibinfo{author}{\bibfnamefont{J.~R.} \bibnamefont{Williams}},
  \bibinfo{author}{\bibfnamefont{T.}~\bibnamefont{Low}},
  \bibinfo{author}{\bibfnamefont{M.~S.} \bibnamefont{Lundstrom}},
  \bibnamefont{and} \bibinfo{author}{\bibfnamefont{C.~M.}
  \bibnamefont{Marcus}}, \bibinfo{note}{arXiv:1008.3704}.

\bibitem[{\citenamefont{Cserti et~al.}(2007)\citenamefont{Cserti, P\'alyi, and
  P\'eterfalvi}}]{csertiprl}
\bibinfo{author}{\bibfnamefont{J.}~\bibnamefont{Cserti}},
  \bibinfo{author}{\bibfnamefont{A.}~\bibnamefont{P\'alyi}}, \bibnamefont{and}
  \bibinfo{author}{\bibfnamefont{C.}~\bibnamefont{P\'eterfalvi}},
  \bibinfo{journal}{Phys. Rev. Lett.} \textbf{\bibinfo{volume}{99}},
  \bibinfo{pages}{246801} (\bibinfo{year}{2007}).

\bibitem[{\citenamefont{Ziegler et~al.}(2009)\citenamefont{Ziegler, D\'ora, and
  Thalmeier}}]{zieglerprb}
\bibinfo{author}{\bibfnamefont{K.}~\bibnamefont{Ziegler}},
  \bibinfo{author}{\bibfnamefont{B.}~\bibnamefont{D\'ora}}, \bibnamefont{and}
  \bibinfo{author}{\bibfnamefont{P.}~\bibnamefont{Thalmeier}},
  \bibinfo{journal}{Phys. Rev. B} \textbf{\bibinfo{volume}{79}},
  \bibinfo{pages}{235431} (\bibinfo{year}{2009}).

\bibitem[{\citenamefont{Shytov et~al.}(2009)\citenamefont{Shytov, Abanin, and
  Levitov}}]{shytovprl}
\bibinfo{author}{\bibfnamefont{A.~V.} \bibnamefont{Shytov}},
  \bibinfo{author}{\bibfnamefont{D.~A.} \bibnamefont{Abanin}},
  \bibnamefont{and} \bibinfo{author}{\bibfnamefont{L.~S.}
  \bibnamefont{Levitov}}, \bibinfo{journal}{Phys. Rev. Lett.}
  \textbf{\bibinfo{volume}{103}}, \bibinfo{pages}{016806}
  (\bibinfo{year}{2009}).

\bibitem[{\citenamefont{Uchoa et~al.}(2009)\citenamefont{Uchoa, Yang, Tsai,
  Peres, and {Castro Neto}}}]{uchoaprl}
\bibinfo{author}{\bibfnamefont{B.}~\bibnamefont{Uchoa}},
  \bibinfo{author}{\bibfnamefont{L.}~\bibnamefont{Yang}},
  \bibinfo{author}{\bibfnamefont{S.~W.} \bibnamefont{Tsai}},
  \bibinfo{author}{\bibfnamefont{N.~M.~R.} \bibnamefont{Peres}},
  \bibnamefont{and} \bibinfo{author}{\bibfnamefont{A.~H.} \bibnamefont{{Castro
  Neto}}}, \bibinfo{journal}{Phys. Rev. Lett.} \textbf{\bibinfo{volume}{103}},
  \bibinfo{pages}{206804} (\bibinfo{year}{2009}).

\bibitem[{\citenamefont{Saremi}(2007)}]{saremiprb}
\bibinfo{author}{\bibfnamefont{S.}~\bibnamefont{Saremi}},
  \bibinfo{journal}{Phys. Rev. B} \textbf{\bibinfo{volume}{76}},
  \bibinfo{pages}{184430} (\bibinfo{year}{2007}).

\bibitem[{\citenamefont{Cheianov and Falko}(2006)}]{cheifalko}
\bibinfo{author}{\bibfnamefont{V.~V.} \bibnamefont{Cheianov}} \bibnamefont{and}
  \bibinfo{author}{\bibfnamefont{V.~I.} \bibnamefont{Falko}},
  \bibinfo{journal}{Phys. Rev. Lett.} \textbf{\bibinfo{volume}{97}},
  \bibinfo{pages}{226801} (\bibinfo{year}{2006}).

\bibitem[{\citenamefont{Cheianov}(2007)}]{cheiepj}
\bibinfo{author}{\bibfnamefont{V.~V.} \bibnamefont{Cheianov}},
  \bibinfo{journal}{Eur. Phys. J. Special Topics}
  \textbf{\bibinfo{volume}{148}}, \bibinfo{pages}{55} (\bibinfo{year}{2007}).

\bibitem[{\citenamefont{Rutter et~al.}(2007)\citenamefont{Rutter, Crain,
  Guisinger, Li, First, and Stroscio}}]{ruttersci}
\bibinfo{author}{\bibfnamefont{G.~M.} \bibnamefont{Rutter}},
  \bibinfo{author}{\bibfnamefont{J.~N.} \bibnamefont{Crain}},
  \bibinfo{author}{\bibfnamefont{N.~P.} \bibnamefont{Guisinger}},
  \bibinfo{author}{\bibfnamefont{T.}~\bibnamefont{Li}},
  \bibinfo{author}{\bibfnamefont{P.~N.} \bibnamefont{First}}, \bibnamefont{and}
  \bibinfo{author}{\bibfnamefont{J.~A.} \bibnamefont{Stroscio}},
  \bibinfo{journal}{Science} \textbf{\bibinfo{volume}{317}},
  \bibinfo{pages}{219} (\bibinfo{year}{2007}).

\bibitem[{\citenamefont{Bena}(2008)}]{benaprl}
\bibinfo{author}{\bibfnamefont{C.}~\bibnamefont{Bena}}, \bibinfo{journal}{Phys.
  Rev. Lett.} \textbf{\bibinfo{volume}{100}}, \bibinfo{pages}{076601}
  (\bibinfo{year}{2008}).

\bibitem[{\citenamefont{Brihuega et~al.}(2008)\citenamefont{Brihuega, Mallet,
  Bena, Bose, Michaelis, Vitali, Varchon, Magaud, Kern, and
  Veuillen}}]{brihuprl}
\bibinfo{author}{\bibfnamefont{I.}~\bibnamefont{Brihuega}},
  \bibinfo{author}{\bibfnamefont{P.}~\bibnamefont{Mallet}},
  \bibinfo{author}{\bibfnamefont{C.}~\bibnamefont{Bena}},
  \bibinfo{author}{\bibfnamefont{S.}~\bibnamefont{Bose}},
  \bibinfo{author}{\bibfnamefont{C.}~\bibnamefont{Michaelis}},
  \bibinfo{author}{\bibfnamefont{L.}~\bibnamefont{Vitali}},
  \bibinfo{author}{\bibfnamefont{F.}~\bibnamefont{Varchon}},
  \bibinfo{author}{\bibfnamefont{L.}~\bibnamefont{Magaud}},
  \bibinfo{author}{\bibfnamefont{K.}~\bibnamefont{Kern}}, \bibnamefont{and}
  \bibinfo{author}{\bibfnamefont{J.~Y.} \bibnamefont{Veuillen}},
  \bibinfo{journal}{Phys. Rev. Lett.} \textbf{\bibinfo{volume}{101}},
  \bibinfo{pages}{206802} (\bibinfo{year}{2008}).

\bibitem[{\citenamefont{Bena}(2009)}]{benaprb}
\bibinfo{author}{\bibfnamefont{C.}~\bibnamefont{Bena}}, \bibinfo{journal}{Phys.
  Rev. B} \textbf{\bibinfo{volume}{79}}, \bibinfo{pages}{125427}
  (\bibinfo{year}{2009}).

\bibitem[{\citenamefont{Simon et~al.}(2009)\citenamefont{Simon, Bena, Vonau,
  Aubel, Nasrallah, Habar, and Peruchetti}}]{simonepjb}
\bibinfo{author}{\bibfnamefont{L.}~\bibnamefont{Simon}},
  \bibinfo{author}{\bibfnamefont{C.}~\bibnamefont{Bena}},
  \bibinfo{author}{\bibfnamefont{F.}~\bibnamefont{Vonau}},
  \bibinfo{author}{\bibfnamefont{D.}~\bibnamefont{Aubel}},
  \bibinfo{author}{\bibfnamefont{H.}~\bibnamefont{Nasrallah}},
  \bibinfo{author}{\bibfnamefont{M.}~\bibnamefont{Habar}}, \bibnamefont{and}
  \bibinfo{author}{\bibfnamefont{J.~C.} \bibnamefont{Peruchetti}},
  \bibinfo{journal}{Eur. Phys. J. B} \textbf{\bibinfo{volume}{69}},
  \bibinfo{pages}{351} (\bibinfo{year}{2009}).

\end{thebibliography}

\end{document}